\documentclass[final,5p,times,twocolumn]{elsarticle}

\usepackage{graphicx}
\usepackage{subfigure}
\usepackage{multirow}

\usepackage[utf8]{inputenc}
\usepackage{amsmath}
\usepackage{amsthm}
\usepackage{amsfonts}
\usepackage{epsfig}
\usepackage{psfrag}
\usepackage{pstricks}
\usepackage{algorithm}
\usepackage{amssymb}
\usepackage{bm}

\usepackage{color}

\definecolor{turquoise}{cmyk}{0.65,0,0.1,0.1}
\definecolor{purple}{rgb}{0.65,0,0.65}
\definecolor{dark_green}{rgb}{0, 0.5, 0}
\definecolor{orange}{rgb}{0.8, 0.2, 0.2}

 \biboptions{comma,sort&compress}

\journal{Elsevier}

\begin{document}

\begin{frontmatter}

\title{Length-optimal tool path planning for freeform surfaces with preferred feed directions}

\author[uman,ubc]{Qiang Zou\corref{cor}}
\ead{john.qiangzou@gmail.com}

\author[uman]{Charlie C. L. Wang}

\author[ubc]{Hsi-Yung Feng}

\cortext[cor]{Corresponding author.}
\address[uman]{The University of Manchester, Manchester, M13 9PL, United Kingdom}
\address[ubc]{The University of British Columbia, Vancouver, BC, V6T 1Z4, Canada}

\begin{abstract}
This paper presents a new method to generate tool paths for machining freeform surfaces represented either as parametric surfaces or as triangular meshes. This method allows for the optimal tradeoff between the preferred feed direction field and the constant scallop height, and yields a minimized overall path length. The optimality is achieved by formulating tool path planning as a Poisson problem that minimizes a simple, quadratic energy. This Poisson formulation considers all tool paths at once, without resorting to any heuristic sampling or initial tool path choosing as in existing methods, and is thus a globally optimal solution. Finding the optimal tool paths amounts to solving a well-conditioned sparse linear system, which is computationally convenient and efficient. Tool paths are represented with an implicit scheme that can completely avoid the challenging topological issues of path singularities and self-intersections seen in previous methods. The presented method has been validated with a series of examples and comparisons.

\end{abstract}


\end{frontmatter}

\section{Introduction}
\label{sec:introduction}
Computer-aided design and manufacturing (CAD/CAM) systems have seen applications in many fields, including automotive, shipbuilding, and aerospace industries. One of the essential elements in a CAD/CAM system is tool path planning, which bridges part geometries designed in CAD with cutting processes controlled in CAM \cite{Altintas2005}. The generated tool paths govern how a three- or five-axis machine tool moves its cutter relative to the part geometry (in particular a freeform surface) during machining. Hence, the quality of tool paths directly impacts the accuracy and efficiency of machining.

Generating high-quality tool paths is, however, no trivial matter due to two major facts. First, scallops are produced between adjacent tool paths, posing the machining accuracy problem \cite{feng2002constant}. Second, the machining strip width at a cutter contact point varies with the feed direction and the cutter orientation, offering the potential for machining efficiency optimization \cite{chiou2002machining}. Ideally, the most efficient tool paths would have a constant scallop height between adjacent tool paths to avoid redundant machining, and meanwhile have the maximum strip width along individual tool paths to attain maximum material removal. However, Kumazawa et al. \cite{kumazawa2015preferred} have shown that, in most cases, tool paths having constant scallop height deviate considerably from those following feed directions of maximum strip width. That is, there is an incompatibility between the constant scallop height and the optimal feed directions, and this poses the challenging problem of finding the best tradeoff between them.

Much previous work (Section \ref{sec:relatedwork}) related to the above problem focused on one of its two sub-problems, either tool paths having a constant scallop height or tool paths following exactly preferred feed directions. (In the literature, it is customary to call optimal feed directions as preferred feed directions; they are thus used interchangeably in this work.) Recently, some work to integrate the two lines of research has been reported \cite{kumazawa2015preferred}. Nevertheless, preferred feed directions were primarily used to assist the generation of constant scallop height tool paths, e.g., help choose the initial tool path, where little can be said about global optimality. In this paper, a new approach is to be presented to address this global tool path optimization problem.

The proposed method (Sections \ref{sec:Methodology} and \ref{sec:Implementation}) expresses the problem stated above in terms of the solution to a Poisson equation (a second-order linear partial differential equation). Unlike many existing methods, we approach the problem using an implicit tool path representation framework: a scalar function defined over the design surface is to be computed, and then tool paths are obtained by extracting appropriate iso-level curves (Fig. \ref{implicit-representation}). Based on this representation, it is found that there is an integral relationship between the optimal tool paths and a vector field collectively representing the two requirements of constant scallop height and preferred feed directions. This allows us to express the tool path optimization problem in terms of computing a scalar function whose gradient best fits the vector field, leading to a standard Poisson problem \cite{crane2017heat}. It should be noted that this work distinguishes vectors that can have any length, and directions that have unit length. As such, previous work can be classified as direction-field-based, while the present work is vector-field-based.

Formulating tool path optimization as a Poisson problem offers a number of advantages. Most notably, it generates the globally optimal tool paths such that the scallop height is kept as constant as possible, the strip width is made as large as possible, and consequently the overall path length can be minimized. It is also conceptually simple and easy to implement as the solution reduces to solving a well-conditioned sparse linear system. In addition, as tool paths are represented as iso-level curves of a scalar function, there is no particular order among them, and thus there is no need to deal with the complex task of determining the initial tool path (a long-standing problem in the tool path planning domain). Another noteworthy benefit of using the implicit representation is that it can automatically handle singularities and self-intersections in tool paths without any tedious, error-prone topological operations \cite{zou2014iso}.

\section{Related work}
\label{sec:relatedwork} 
Tool path planning is an extensively studied problem in CAD/CAM, and many methods have been reported \cite{lasemi2010recent}. Among those methods, the categories of interest to this work are the constant scallop height (or iso-scallop) paradigm and the preferred feed direction paradigm. The iso-scallop paradigm was proposed as an improvement to the previous iso-parametric and iso-planar paradigms such that redundant machining observed in those two paradigms can be avoided. This paradigm was initialized by Suresh and Yang \cite{suresh1994constant} and improved in computation accuracy in Refs. \cite{sarma1997geometry, feng2002constant, kim2007constant}, efficiency in Refs. \cite{koren1996efficient, tournier2002surface}, and applicability in Refs. \cite{Lee1998Nonisoparametric, li2004efficient, wen2017cutter}. Although presented in different forms, they have a common idea: sequentially offset the current tool path to attain the next tool path while keeping the resulting scallop height the same as a specified limit. In this way, no redundant machining exists, and a shorter overall path length than iso-parametric and iso-planar tool paths is attained.

Instead of eliminating redundant machining between adjacent tool paths, one can reduce the overall path length through maximizing the machining strip width along individual tool paths. The fundamental principle behind this method is that the design surface's area approximately equals the integrated strip width along tool paths if there is no overlapping between adjacent machining strips (a.k.a. no redundant machining). Mathematically, $ Area = \sum_{i}{w_i d_i}$ where the sum is taken over all cutter contact points $i$, and $w_i, d_i$ are the corresponding strip width and forward-step. Clearly, increasing each $w_i$ can decrease the overall path length $\sum_{i}{d_i}$. This line of research was pioneered by Chiou and Lee \cite{chiou2002machining} and improved in a series of papers \cite{Chen2004Principle, anotaipaiboon2005tool, moodleah2016five, Sun2017, ma2020toolpath}. Basically, they sought to align tool paths with a preferred feed direction field of maximum strip width.

Recently, Kumazawa et al. \cite{kumazawa2012generating, kumazawa2015preferred} have, however, shown that tool paths following preferred feed directions do not necessarily mean a minimized, or even shorter, overall path length, because the condition of no redundant machining cannot be satisfied in general. Specifically, if preferred feed directions are demanded, the resulting tool paths deviate considerably from being iso-scallop, and vice versa. To mitigate this problem, they suggested a hybrid method: segment the design surface into patches based on the preferred feed directions and then generate iso-scallop tool paths within each patch. A similar idea was also presented in \cite{liu2015tool}. Su et al. \cite{Su2020Initial} further proposed an optimization procedure to choose a proper initial tool path for generating iso-scallop tool paths within each patch. Using this hybrid strategy, improved alignment between iso-scallop tool paths and preferred feed directions were demonstrated. Nevertheless, this way of working can only achieve sub-optimal tool paths. Big misalignment still exists in the generated tool paths, refer to, for example, Fig. 12 in Ref. \cite{kumazawa2015preferred}.

To date, substantial progress has been made in understanding and addressing the problem of the optimal tradeoff between the constant scallop height and the preferred feed direction field with the goal of minimizing the overall path length. Nevertheless, a globally optimal solution has yet been made available. This work follows this research direction but uses a new way to approach the problem. It outlines the basic ideas/formulations in Section \ref{sec:Methodology} and elaborates them with implementation details in Section \ref{sec:Implementation}. Validation of the method using a series of examples and comparisons are found in Section \ref{sec:results}, followed by conclusions in Section \ref{sec:conclusion}.

\section{Methodology}
\label{sec:Methodology}
As already noted, tool paths would have a minimized overall length if they have a constant scallop height and follow preferred feed directions, but in general these two objectives cannot be satisfied concurrently. Minimizing the overall path length thus lies in finding the closest satisfaction of the two objectives, i.e., the globally optimal tradeoff between the constant scallop height and the preferred feed direction field. Leaning towards either side, as in existing methods, could only give sub-optimal tool paths.

The above optimal alignment problem can be stated more precisely as follows. Given a surface $S \subseteq \mathbb{R}^{3}$ to be machined, a preferred feed direction field $D \subseteq \mathbb{S}^2$ on tangent planes of $S$, and a scallop height constraint $h \in \mathbb{R}^+$, find tool paths $\{C_i\}_{i=1}^n, C_i \subseteq S$ such that the following two error terms are minimized: (1) the error between $D$ and the direction field of tool path tangents $\{C_i^\prime (p)/\|C_i^\prime (p)\|\}$ where $p \in C_i$ and $\|\cdot\|$ denotes a vector's magnitude; and (2) the error between $h$ and the actual scallop height between adjacent paths $C_i, C_{i+1}$.

In the general form described, the above problem finds challenges in how to define the two error terms, and more importantly how to express them under a common framework for simultaneous optimization. To solve these challenges, we provide a new tool path generation method that consists primarily of two steps: (1) formulate the two error terms as a vector field on tangent planes of $S$; and (2) convert the error minimization problem into a Poisson problem. In these two steps, there is an underlying technique called implicit tool path representation. In the next few subsections, we begin with an introduction to this representation scheme and then describe the two steps.

\subsection{Implicit tool path representation}
\label{sec:tool-path-representation}
\begin{figure}[t]
	\centering
	\includegraphics[]{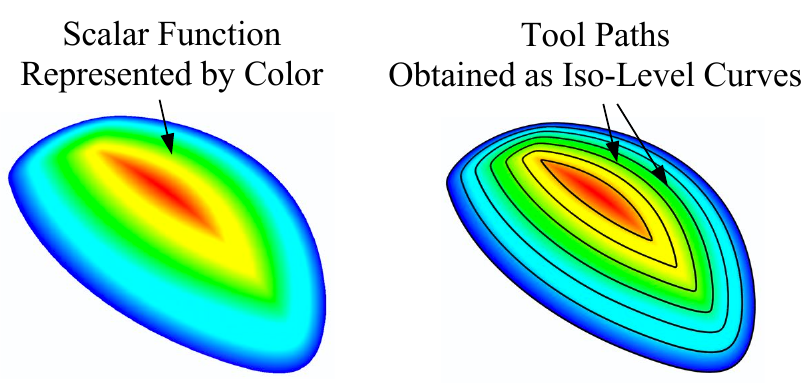}
    \caption{Illustration of implicit tool path representation.}
	\label{implicit-representation}
\end{figure}

The proposed method employs an implicit scheme to represent tool paths. Consider a surface $S$ and a scalar function $\varphi: S\to \mathbb{R}$ defined over it. We define the surface curves that are mapped to a set of values $\{l_i\}_{i=1}^n$ as tool paths. Fig. \ref{implicit-representation} shows one such example. If we change $\varphi$, the shape of tool paths varies accordingly. Geometric properties like tool path tangent directions and path intervals can also be expressed in terms of $\varphi$ without evaluating the corresponding tool paths \cite{zou2014iso}. As a result, tool path optimization transforms into scalar function optimization.

\subsection{Optimal vector field construction}
\label{sec:vector-field-construction}
\begin{figure}[b]
	\centering
	\includegraphics[]{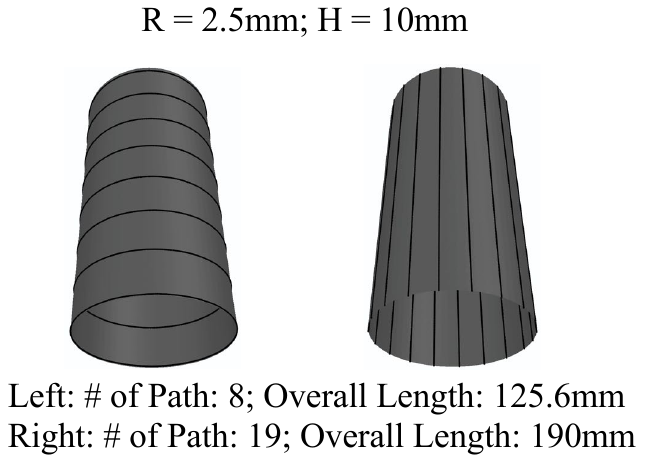}
    \caption{Tool paths generated for a cylinder with scallop height 0.05mm and cutter radius 5mm.}
	\label{cylinder-paths}
\end{figure}

\begin{figure*}[t]
	\centering
	\includegraphics[]{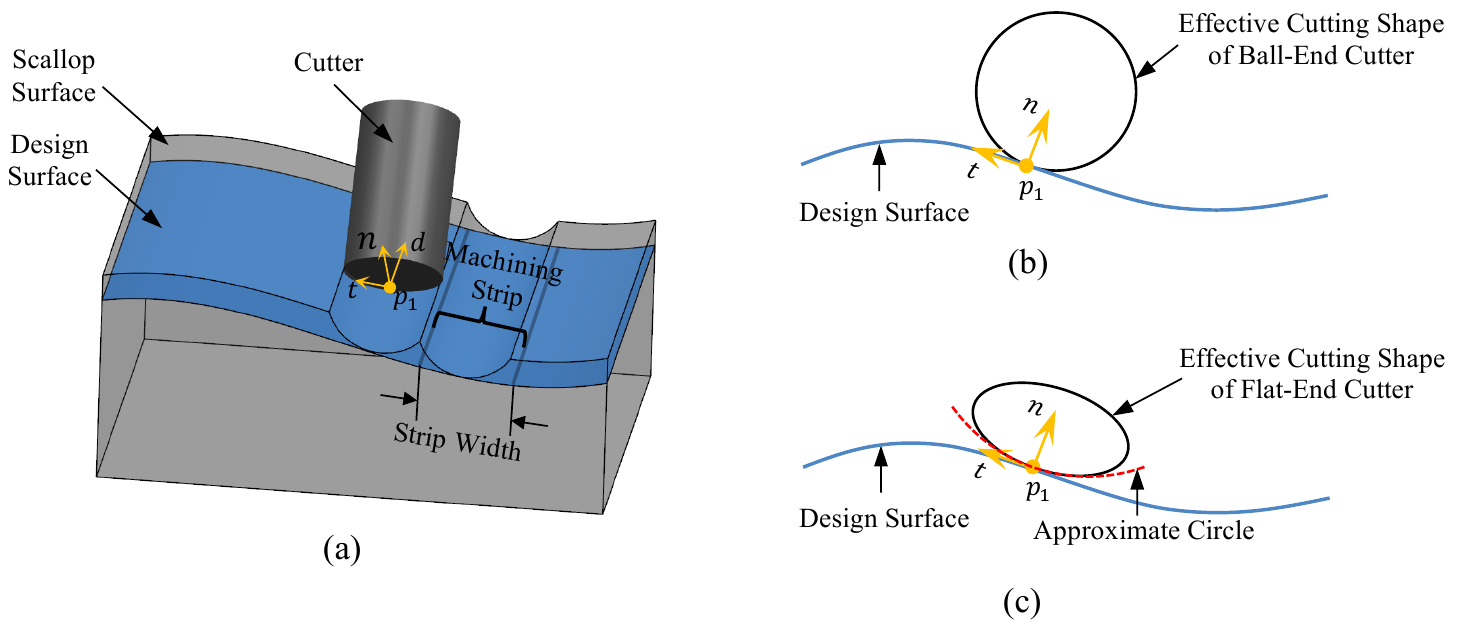}
    \caption{Illustration of machining strip (a) and effective cutting shapes (b) and (c).}
	\label{path-parameters}
\end{figure*}

\noindent \textbf{Preferred feed direction} \quad The fundamental principle of the preferred feed direction method is that the choice of feed directions can make a difference to the overall tool path length. As can be seen from Fig.~\ref{cylinder-paths}, the length of tool paths in the axial direction is $>50\%$ more than those in the circumferential direction. This relationship between feed directions and machining efficiency appears to be first formalized by Lee and Ji \cite{chiou2002machining}, yielding the notion of machining strip width. For this notion, there are various definitions in the literature \cite{li2004efficient}. The one used here is: the machining strip is a strip-like region on the design surface that is right underneath the cutter swept envelope (below the scallop surface) during a cutting pass, and the machining strip width refers to the maximum of the strip's reach in the direction perpendicular to the feed direction at a cutter contact point (Fig.~\ref{path-parameters}a). 

The machining strip width is often evaluated with the help of a 2D sectional geometry, together with the notion of effective cutting shapes (Fig.~\ref{path-parameters}b and \ref{path-parameters}c). The section plane is positioned at a cutter contact point $p_1$ and perpendicular to the feed direction $d$. The effective cutting shape is the projected geometry from the cutter profile (without the cutter’s shank) onto this plane. Depending on the cutter type being used, various effective cutting shapes can be produced. For example, it produces a circular arc in ball-end milling (Fig.~\ref{path-parameters}b) and an elliptical arc in flat-end milling (Fig.~\ref{path-parameters}c). Variety of this kind complicates the strip width calculation. Fortunately, according to \cite{lo1999efficient}, we can instead use a second-order approximation scheme: approximate the neighborhood of an effective cutting shape at $p_1$ using its osculating circle (Fig.~\ref{path-parameters}c). In this way, strip width calculations can be carried out as if effective cutting shapes were all circular arcs.

With the effective cutting shapes in place, the strip width can be easily evaluated using existing algorithms, say \cite{barakchi2010effective,kumazawa2015preferred}. When a cutter changes its feed direction and cutter orientation, the corresponding strip width also changes. This means that a preferred feed direction can be computed at each cutter contact point such that the strip width is maximized. Several efficient algorithms to find the preferred feed direction exist in the literature, see \cite{barakchi2010effective} for a good example. We omit the details here since we do not consider generating preferred feed directions as new results. It is supposed in this work that preferred feed directions have been pre-assigned.

To align iso-level curves of a scalar function $\varphi$ with a preferred feed direction field $D$, we should let tangent directions of the iso-level curves be in line with the directions in $D$. As known, the tangent direction of an iso-level curve at point $p$ is perpendicular to the gradient $\nabla \varphi$ at $p$ ($\nabla \varphi$ is a vector in the tangent plane at $p$) \cite{Carmo76}. The alignment can thus be done by making the gradient field $\nabla \varphi$ perpendicular to $D$, i.e., minimizing the following alignment energy:
\begin{equation}\label{equ-direction-align} 
  E_{Align}(\varphi) = \int_{\mathcal{S}} \left\|D \cdot \nabla \varphi\right\|^2,
\end{equation}
where $``\cdot"$ denotes the inner product.

\begin{figure*}[htbp]
	\centering
	\includegraphics[]{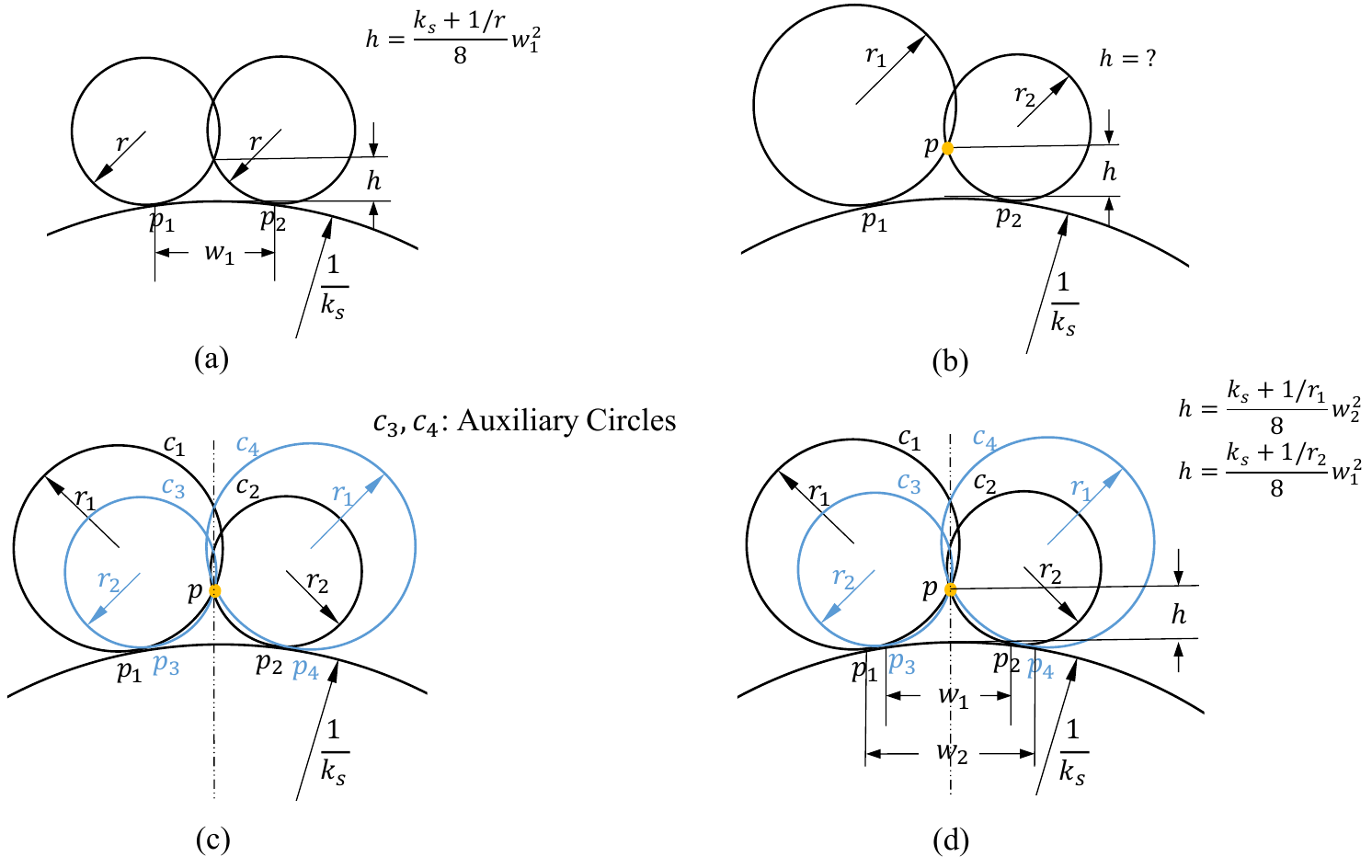}
    \caption{Illustration of scallop geometry: (a) ball-end milling; (b) general milling; (c) auxiliary circles; and (d) scallop height calculation.}
	\label{fig:scallop-geometry}
\end{figure*}

\noindent \textbf{Constant scallop height} \quad The above alignment energy $E_{Align}$ dictates the directions of $\nabla \varphi$. Next, we show how the magnitudes of $\nabla \varphi$ can be associated with the scallop height. The scallop is the uncut volume left between a pair of adjacent cutting passes, and the scallop height refers to the maximum of the uncut volume’s thickness, locally. In ball-end milling, the scallop height for a cutter with radius $r$ is given by \cite{koren1996efficient}:
\begin{equation}\label{eq-scallop-height-ball}
	h=\frac{k_s+\frac{1}{r}}{8}{{{\left\| p_2 - p_1 \right\|}^{2}}} + {O}\left( {{\left\| {{p}_{2}}-{{p}_{1}} \right\|}^{3}}\right),
\end{equation}
where $k_s$ is the normal curvature in the direction perpendicular to the feed direction at $p_1$, and $p_2$ is the cutter contact point at the adjacent tool path corresponding to $p_1$, as shown in Fig.~\ref{fig:scallop-geometry}a. $\left\| p_2 - p_1 \right\|$ measures the distance between the two contact points, which is commonly referred to as the side-step. In the above equation, $k_s$ is positive if the design surface is convex, and negative if concave.

If using other cutter types, the cutting radiuses at $p_1, p_2$ become different (Fig.~\ref{fig:scallop-geometry}b) due to the possible variation of effective cutting shapes at the two points.\footnote{The notion of effective cutting circles would result in a poor estimate of the scallop height, if the feed directions at $p_1, p_2$ differ significantly \cite{feng2002constant,barakchi2010effective}. This work thus assumes a smooth variation of feed directions.} To extend Eq.~\ref{eq-scallop-height-ball} to handle such situations, we introduce two auxiliary circles as shown in Fig.~\ref{fig:scallop-geometry}c: circle $c_3$ with radius $r_2$ is constructed to be tangential to the bottom circle and to pass the scallop point $p$ (indicated by orange dots), and similarly for the other circle $c_4$. It is straightforward to see that circles $c_1, c_4$ and circles $c_2, c_3$ yield the same scallop height as that by circles $c_1, c_2$, and the scallop height is given by:
\begin{equation}\label{eq-scallop-heights}
	\begin{aligned} 
	h &= \frac{k_s+\frac{1}{r_1}}{8}{{{\left\| p_4 - p_1 \right\|}^{2}}} + {O}\left( {{\left\| {{p}_{4}}-{{p}_{1}} \right\|}^{3}}\right), \\
	h &= \frac{k_s+\frac{1}{r_2}}{8}{{{\left\| p_2 - p_3 \right\|}^{2}}} + {O}\left( {{\left\| {{p}_{2}}-{{p}_{3}} \right\|}^{3}}\right).
	\end{aligned}
\end{equation}
Also, circles $c_1, c_4$ and circles $c_2, c_3$ have a line of symmetry as shown in Fig.~\ref{fig:scallop-geometry}d, leading to the following nice relationship between side-steps:
\begin{equation}\label{eq-side-steps}
	\left\|p_2 - p_1\right\| = \frac{\left\|p_4 - p_1\right\| + \left\|p_2 - p_3\right\|}{2} + {O}\left( \left\| {{p}_{2}}-{{p}_{1}} \right\|^3\right).
\end{equation}
(This equation can be understood by approximating the side-step $\|p_i - p_j\|$ with the the corresponding arc bounded by $p_i, p_j$, and this is generally acceptable because in practice $\|p_i - p_j\|$ is much smaller than the radius $1/k_s$.) By substituting Eq.~\ref{eq-scallop-heights} into Eq.~\ref{eq-side-steps}, we have the following general relationship between the scallop height and the side step:
\begin{equation}\label{eq-scallop-height-general}
	\left\|p_2 - p_1\right\| = \sqrt{h} \left(\sqrt{\frac{2}{k_s+1/r_1}} + \sqrt{\frac{2}{k_s+1/r_2}}\right) + {O}\left( \left\| {{p}_{2}}-{{p}_{1}} \right\|^3\right).
\end{equation}
It should be noted that this equation is only a second-order approximation to the actual scallop height. We will analyze its approximation error in Section \ref{sec:results}.

Using the implicit tool path representation scheme, the side-step $\left\|p_2 - p_1\right\|$ in Eq.~\ref{eq-scallop-height-general} can be translated into the level increment of $\varphi$. Let iso-level curves $C_1 = \left\{ p\in S\ |\ \varphi (p)={{l}_{1}} \right\}$ and $C_2 = \left\{ p\in S\ |\ \varphi (p)={{l}_{2}}\right\}$ be the two tool paths passing points $p_1$ and $p_{2}$. Taylor expansion states that:
\begin{equation}\label{equ-Taylor-expansion0}
	{{l}_{2}}-{{l}_{1}}={{\left( \nabla \varphi  \right)}^{T}}\left( {{p}_{2}}-{{p}_{1}} \right)+{O}\left( {{\left\| {{p}_{2}}-{{p}_{1}} \right\|}^{2}} \right).
\end{equation}
As $\nabla \varphi$ is a vector perpendicular to the tangent direction of $C_1$ at $p_1$, we can rewrite the above equation as:
\begin{equation}\label{equ-Taylor-expansion1}
\left| {{l}_{2}}-{{l}_{1}} \right| = \left\| \nabla \varphi  \right\|\cdot \left\| {{p}_{2}}-{{p}_{1}} \right\|+{O}\left( {{\left\| {{p}_{2}}-{{p}_{1}} \right\|}^{2}} \right),
\end{equation}
Here $\left| \cdot \right|$ returns the absolute value. Equivalently,
\begin{equation}\label{equ-Taylor-expansion}
\left\| \nabla \varphi \right\| = \lim_{\left\| {{p}_{2}}-{{p}_{1}} \right\| \rightarrow 0} \frac{\left| {{l}_{2}}-{{l}_{1}} \right|}{\left\| {{p}_{2}}-{{p}_{1}} \right\|}.
\end{equation}
If we further endow the increment $\left\| {{l}_{2}}-{{l}_{1}} \right\|$ with a physical meaning---equal to the square root of scallop height---Eq.~\ref{equ-Taylor-expansion} and Eq.~\ref{eq-scallop-height-general} can be combined into:
\begin{equation}\label{equ-magnitude-condition}
	\begin{split}
	\left\| \nabla \varphi  \right\| &= \lim_{\left\| {{p}_{2}}-{{p}_{1}} \right\| \rightarrow 0} \frac{1}{\sqrt{\frac{2}{k_s+1/r_1}} + \sqrt{\frac{2}{k_s+1/r_2}}}\\
		&=	\frac{1}{2\sqrt{\frac{2}{k_s+1/r_1}}} \quad  (\text{note:\ }r_2 = r_1 \ \text{when} \ p_2 \rightarrow p_1) \\
		&= \sqrt{\frac{k_s+1/r_1}{8}}
	\end{split}
\end{equation}
Under this equation, two iso-level curves with increment $\sqrt{h}$ are two tool paths of constant scallop height $h$. As a result, generating iso-scallop tool paths can be converted into satisfying this equation, or minimizing the following scallop energy:
\begin{equation}\label{equ-magnitude-align} 
	E_{Scallop}(\varphi) = \int_{\mathcal{S}} \left(\left\|\nabla \varphi\right\| - \sqrt{\frac{k_s+1/r_1}{8}} \right)^2.
\end{equation}

\noindent \textbf{Optimal vector field} \quad Having formed the alignment energy (\ref{equ-direction-align}) and the scallop energy (\ref{equ-magnitude-align}), we can attain the globally optimal tool paths regarding the constant scallop height and the preferred feed direction field via solving the following optimization problem:
\begin{equation}\label{equ-global-optimization0} 
	\min_{\varphi} E_{Align}(\varphi) + E_{Scallop}(\varphi)
\end{equation}
However, this problem is hard to solve due to the high nonlinearity. To resolve this issue, we rewrite the alignment energy in the following way. Instead of using the inner product, we can equivalently express the energy in terms of a cross product and a $90^\circ$ rotated version of $D$:
\begin{equation}\label{equ-direction-align-rot} 
	E_{Align}(\varphi) = \int_{\mathcal{S}} \left\|D^{90^\circ} \times \nabla \varphi\right\|^2,
\end{equation}
where $D^{90^\circ}$ rotates $D$ by $90^\circ$ about the surface normal, and $``\times"$ denotes the cross product. As such, the alignment energy encourages $\nabla \varphi$ to point in the direction of $D^{90^\circ}$, and the scallop energy encourages $\nabla \varphi$ to further have magnitudes defined by Eq.~\ref{equ-magnitude-condition}. That being said, the two energies, when working together, push $\nabla \varphi$ towards the following vector field:
\begin{equation}\label{equ-vector-field} 
	\begin{aligned} 
		V.\text{direction} &\leftarrow D^{90^\circ}, \\
		V.\text{magnitude} &\leftarrow \sqrt{\frac{k_s+1/r_1}{8}}.
	\end{aligned}
\end{equation}
Then, optimal tool paths can be attained by solving the following linear least squares problem:
\begin{equation}\label{equ-global-optimization} 
	\min_{\varphi} \int_{\mathcal{S}} \left\|~\nabla \varphi - V~\right\|^2.
\end{equation}

\subsection{Optimal tool path generation}
\label{sec:optimaltoolpath}
The optimization problem (\ref{equ-global-optimization}) is a well studied problem in computer graphics and computational mechanics. A standard procedure to deal with it is to solve its corresponding Euler-Lagrange equation:
\begin{equation}\label{equ-Euler-Lagrange equation}
	\Delta \varphi = \nabla \cdot V,
\end{equation}
where $\Delta$ is the Laplacian operator, and $\nabla\cdot$ the divergence operator \cite{botsch2010polygon}. This is a standard Poisson equation---a second-order linear partial differential equation. In the numerical setting, it becomes a sparse linear system of equations that can be solved efficiently using existing methods, see details in Section \ref{sec:Implementation}.

Assume that Eq.~\ref{equ-Euler-Lagrange equation} has been successfully solved, and that the optimal scalar function $\varphi$ has been made available. Then, the last missing piece in generating optimal tool paths is to extract tool paths based on $\varphi$. That is, we want to determine level values $\{l_i\}_{i=1}^n$ whose corresponding iso-level curves are to be used as tool paths. This can be done by a modification of the method to calculate path intervals for iso-parametric tool paths. The specific procedures are: (1) a certain number of points are sampled from a iso-level curve $C_i$; (2) for each point, the level increment $\left| {{l}_{i+1}}-{{l}_{i}} \right|$ is computed with respect to the given scallop height $h$; and (3) the smallest level increment is chosen as the level increment between $C_i$ and its next path $C_{i+1}$. Having found $l_{i+1}$, the corresponding iso-level curve on the surface can be attained with the marching triangle algorithm presented in Ref. \cite{zou2014iso}.

In summary, the overall method consists of three major steps:
\begin{enumerate}
  \item Construct the optimal vector field $V$ by using (\ref{equ-vector-field});
  \item Solve the Poisson equation (\ref{equ-Euler-Lagrange equation}); and
  \item Extract iso-level curves from the optimal scalar function $\varphi$.
\end{enumerate}
The next section will provide more technical details about these steps. Before closing this section, a few notes about the proposed method's extensibility are provided. The descriptions presented so far are all related to feed directions of maximum strip width, but the described principles are readily applicable to other feed direction fields such as the kinematics-derived direction field investigated in Ref. \cite{kim2002toolpath}. Also, we can adapt the proposed framework of optimization to include the generation of smooth tool paths, iso-scallop tool paths, and tool paths following exactly preferred directions, see \ref{append-extend}.

\section{Implementation}
\label{sec:Implementation}
This section provides implementation details of the method outlined in the previous section, concerning numerical solutions to the Poisson equation, the consistency issue of feed directions, the singular region issue where preferred feed directions are ill-defined, and the issue of non-smooth preferred feed directions. These issues (except for the first) are not only of interest to this work but also important to previous research studies built upon preferred feed directions. There have, however, been limited details given by those work.

\subsection{Numerical solutions}
The Poisson equation (\ref{equ-Euler-Lagrange equation}) may be numerically solved using the finite element method or the iso-geometric analysis method; both of them are sound methods and able to provide satisfactory results. This work employs the finite element method, and this makes the proposed method applicable to both parametric surfaces (with an additional meshing step \cite{botsch2010polygon}) and triangular mesh surfaces. 

\begin{figure}[htbp]
	\centering
	\includegraphics[]{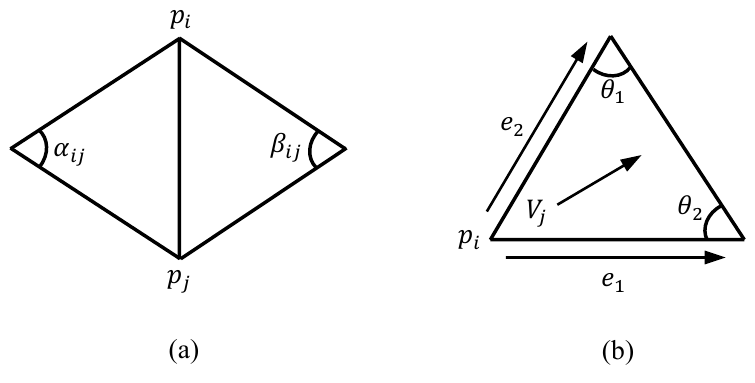}
	\caption{Triangles for computing Laplaican (a) and divergence (b).}
	\label{fig-Differential-Operators}
\end{figure}

The Poisson equation involves two linear differential operators: Laplacian $\Delta$ and divergence $\nabla\cdot$. For mesh surfaces, their standard definitions are as follows \cite{botsch2010polygon}. The Laplacian of a scalar function $\varphi$ gives another scalar function whose value at a mesh vertex $p_i$ is given by:
\begin{equation}\label{equ-Laplacian}
	(\Delta \varphi)_i = \frac{1}{2A_i}\sum\limits_{j} \left( \cot \alpha_{ij} + \cot \beta_{ij} \right)\left( \varphi_j - \varphi_i \right),
\end{equation}
where $A_i$ is the Voronoi area of $p_i$, and the sum is taken over all neighboring vertices $p_j$, and $\alpha_{ij}, \beta_{ij}$ are shown in Fig~\ref{fig-Differential-Operators}a. The divergence operator takes in a vector field $V$ and gives back a scalar field with value at a vertex $p_i$:
\begin{equation}\label{equ-div}
	(\nabla\cdot V)_i = \frac{1}{2A_i}\sum\limits_{k}\cot\theta_1(e_{1} \cdot V_j) + \cot\theta_{2}(e_{2} \cdot V_j),
\end{equation}
where the sum is taken over all incident triangles $k$ each with vector $V_j$, and $\theta_1, \theta_2, e_1, e_2$ are shown in Fig.~\ref{fig-Differential-Operators}b. With Eq.~\ref{equ-Laplacian} and \ref{equ-div}, Eq.~\ref{equ-Euler-Lagrange equation} becomes a sparse system of linear equations, which can be effectively solved with existing linear algebra libraries such as Eigen.

\subsection{Consistent preferred feed directions}
It is important to note that, at a cutter contact point, the machining strip widths are the same for two opposite feed directions, if gouging is not a concern \cite{kumazawa2015preferred}. For this reason, there could be two feasible optimal feed directions (opposite to each other) at one cutter contact point. Cutting passes in either direction are applicable as long as all the feed directions are consistently oriented. If the condition does not hold, some directions should be flipped. To do so, we start out with an arbitrarily chosen seed triangle in the mesh, then propagate the direction to its neighboring triangles, and repeat until the mesh is covered. This is conceptually similar to the breadth-first search (BFS). When there are triangles with unique feed directions due to, for example, gouge avoiding, all such triangles will be used as seed triangles in the propagation.

\begin{figure}[htbp]
	\centering
	\includegraphics[]{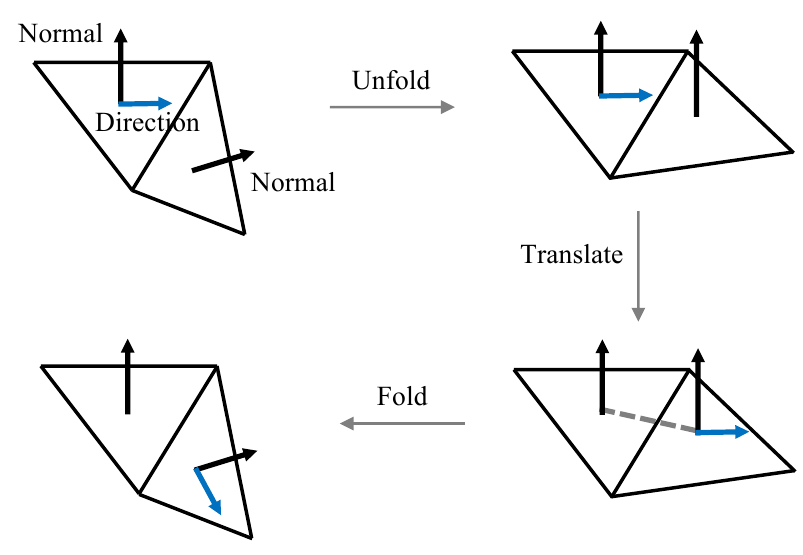}
	\caption{Transport a direction from a triangle to its neighboring triangle.}
	\label{fig-unfold-triangles}
\end{figure}

A special case is that there could be singular regions (e.g., flat regions) where any feed direction at a point gives the same largest strip width, and thus no preferred feed direction can be determined. For such regions, we need to extrapolate direction information from neighboring well-defined regions. To do so, we slightly modify the above propagation scheme: instead of flipping directions, directions in ill-defined regions are copied from parent triangles to children triangles during the propagation. As two neighboring triangles are generally not coplanar, the copying cannot be done by simply translating directions from parent triangles to children triangles, but consists of the three steps as depicted in Fig.~\ref{fig-unfold-triangles}. (It should be noted that the steps here do not factor in the holonomy \cite{Crane2013DGP} in vector transportation, possibly resulting in slightly non-smooth feed directions. To neatly solve this issue, the notion of trivial connection needs to be used, but this requires significant introduction overheads of differential geometry. This work opts for post-processing of the directions using Laplacian smoothing \cite{botsch2010polygon}.)

\subsection{Surface segmentation}
\label{sec:surface-segment}
\begin{figure}[htbp]
	\centering
	\includegraphics[]{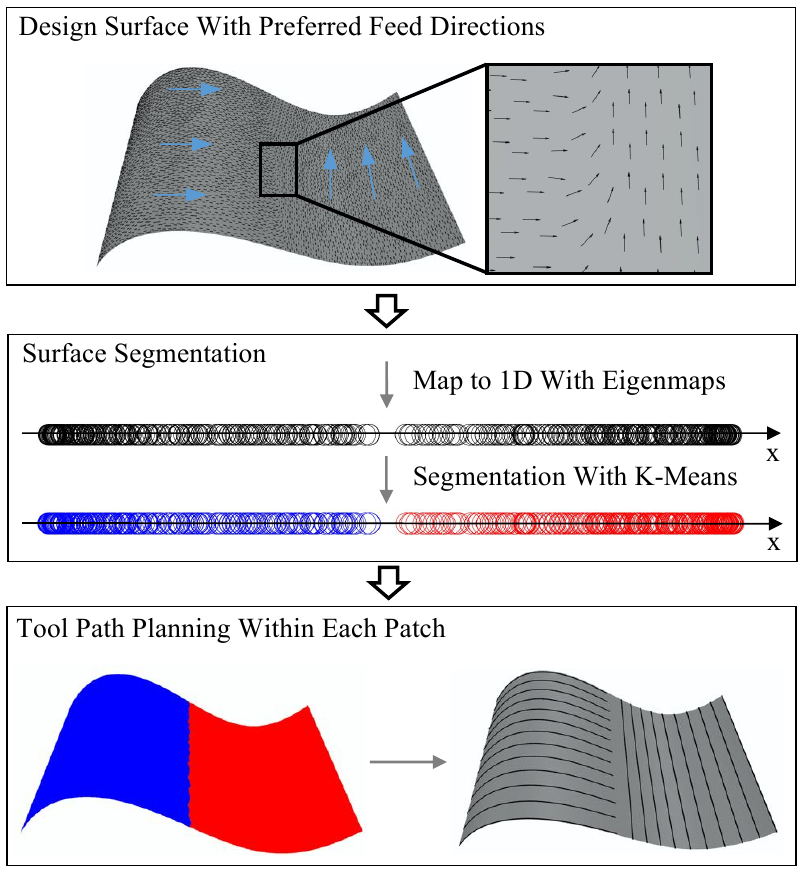}
	\caption{Surface segmentation using combination of Eigenmaps and K-Means (circles on the x-axis: mapped points).}
	\label{fig-surface-segmentation}
\end{figure}

It was assumed in Section \ref{sec:Methodology} (as well as in most previous work) that feed directions vary smoothly over the design surface. However, this is not always the case, e.g., the freeform surface shown in the top figure of Fig.~\ref{fig-surface-segmentation}. At points where the convex region meets the concave region, there are abrupt direction changes from rightwards to upwards. Another source of such abrupt changes is due to a special case in consistent feed direction generation. If there are multiple seed triangles and some of them have inconsistent directions, then abrupt changes could occur at borders between the BFS-grown regions. Abrupt changes could also be caused by degenerate points \cite{kumazawa2015preferred} in a preferred feed direction field. Nevertheless, degenerate points often have a very low number (e.g., two or three), and are sparsely distributed over the surface. As a result, they have a very limited impact on the global optimization method presented. Thus, no special procedure is needed to handle them for the proposed method, although they are the primary concern in previous work.

To address the problem stated above, the design surface is segmented into distinct patches within which there are no abrupt changes among feed directions. Basically, we need to cut along points where abrupt changes occur. To do so, we employ the following idea: by moving neighboring points with low direction variations close to each other and pushing those with high direction variations away from each other, the feed direction dissimilarity is magnified, and then the cuts would emerge by themselves. The implementation of this idea consists of three steps. First, we define a metric to measure the closeness of feed directions at neighboring points:
\begin{equation}
	\text{exp}\left(-\frac{(1 - d_1 \cdot d_2)^2}{2\sigma^2}\right),
\end{equation}
where $d_1, d_2$ are the directions at two neighboring points in question, and $\sigma$ is a free parameter that can be set to 0.67 $(\approx 2 / 3, \text{and max}(1 - d_1 \cdot d_2) = 2)$ in this work. 

At step 2, we use the metric to compute a mapping from the design surface to a line so that two neighboring points with a low direction variation stay as close together as possible. This mapping implements the moving/pushing part of the idea stated above. To find such a mapping, the Laplacian Eigenmaps method \cite{belkin2003laplacian} developed by the pattern recognition community is a good fit. This method consists primarily of two computationally efficient steps: construct a Laplacian using the above metric as the weight function, and then compute the eigenvector corresponding to the smallest non-zero eigenvalue. Entries of the eigenvector are positions of the points mapped into the line.

Having formed the mapping, surface points transform into clusters of 1D points, as shown by the mid figure in Fig.~\ref{fig-surface-segmentation}. At step 3, we extract these clusters from the 1D points, using the K-Means clustering method \cite{jain2010data}. This method is conceptually very simple and able to group the 1D points into k clusters in which each point belongs to the cluster with the nearest mean. Geometrically, this means that close points are clustered, and points at a distance are discriminated. After the clustering, we send back the partition information to the design surface in 3D, resulting in distinct patches. For each patch segmented, the method presented previously is then applied to generate tool paths to cover the patch, as shown by the bottom figure in Fig.~\ref{fig-surface-segmentation}.

It should be noted that the implementation details presented in this section are only viable methods, not necessarily mean the only or the best ones. For example, one can use alternative surface segmentation methods. Nevertheless, the main result, i.e., Eq.~\ref{equ-global-optimization}, of this work remains unchanged.

\section{Results and discussion}
\label{sec:results}
Three case studies, based on a C++ implementation and a 2.4 GHz Intel Core i5 with 8G memory, are to be presented to demonstrate the effectiveness of the proposed method. Case study 1 considered simple situation where no segmentation is needed; Case study 2 analyzed a comprehensive situation where the surface has to be segmented into multiple patches; Case study 3 involved a standard saddle surface, which was used to carry out error analysis. Case studies 1 and 2 also presents comparisons with the classic iso-scallop method by Feng and Li \cite{feng2002constant} and the state-of-the-art method by Su et al. \cite{Su2020Initial}. The comparison results are summarized in Table \ref{comparison-table}.

\subsection{Case studies}
\begin{figure*}[t]
	\centering
	\includegraphics[]{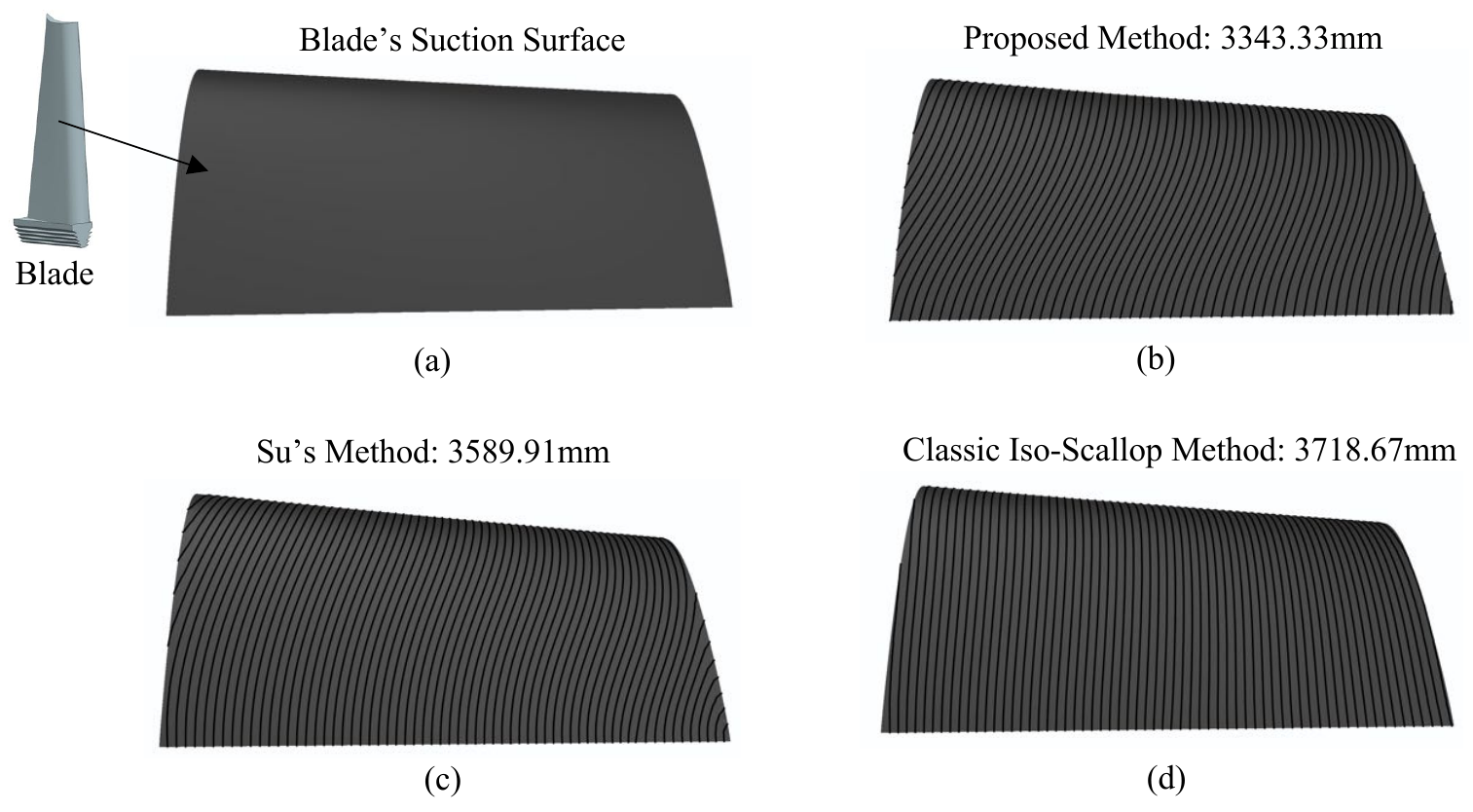}
	\caption{Tool path generation for a blade model (a, b) and comparisons with Su’s method \cite{Su2020Initial} (c) and the classic Feng's method \cite{feng2002constant} (d).}
	\label{fig-blade-result}
\end{figure*}

\begin{figure*}[t]
	\centering
	\includegraphics[]{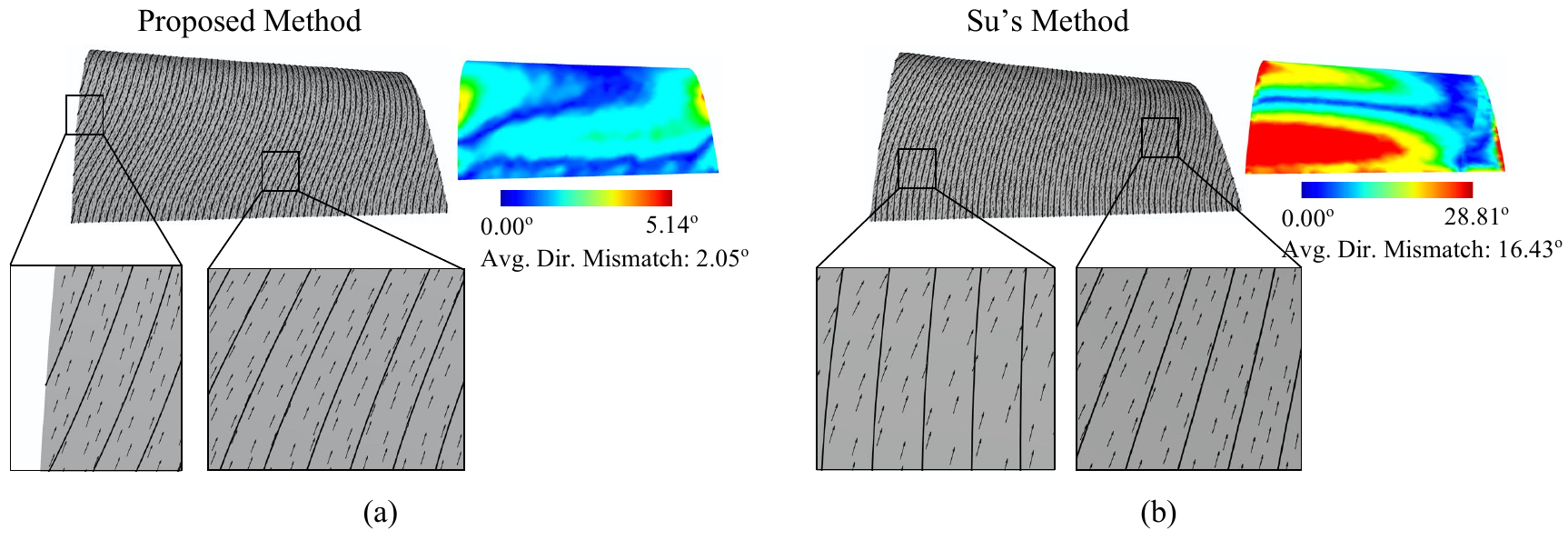}
	\caption{Direction alignment analysis of tool paths generated by the proposed method (a) and Su’s method (b).}
	\label{fig-blade-analysis}
\end{figure*}

Case study 1 considered a blade's suction surface (Fig.~\ref{fig-blade-result}a), which is based on real data downloaded from the GrabCAD part library (https://grabcad.com/library). The generated tool paths are shown in Figs.~\ref{fig-blade-result}b, \ref{fig-blade-result}c and \ref{fig-blade-result}d. A flat-end mill with radius 2mm was chosen, the scallop height constraint was set as 0.1mm, and a constant tilt angle $0^\circ$ and inclination angle $30^\circ$ were used. The alignment of the generated tool paths with the preferred feed directions is shown in Fig.~\ref{fig-blade-analysis}. The mismatch error distributions are also given, together with zoom-in views. 

\begin{figure*}[t]
	\centering
	\includegraphics[]{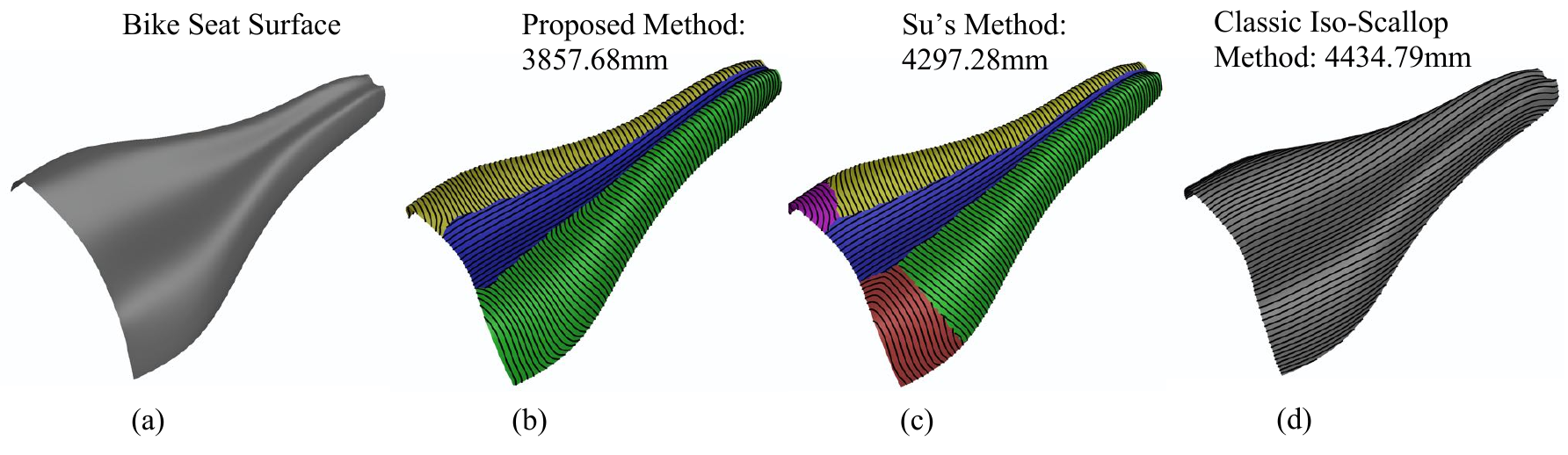}
	\caption{Tool path generation for a seat model (a, b) and comparisons with Su’s method (c) and the classic Feng's method (d).}
	\label{fig-seat-result}
\end{figure*}

\begin{figure}[t]
	\centering
	\includegraphics[]{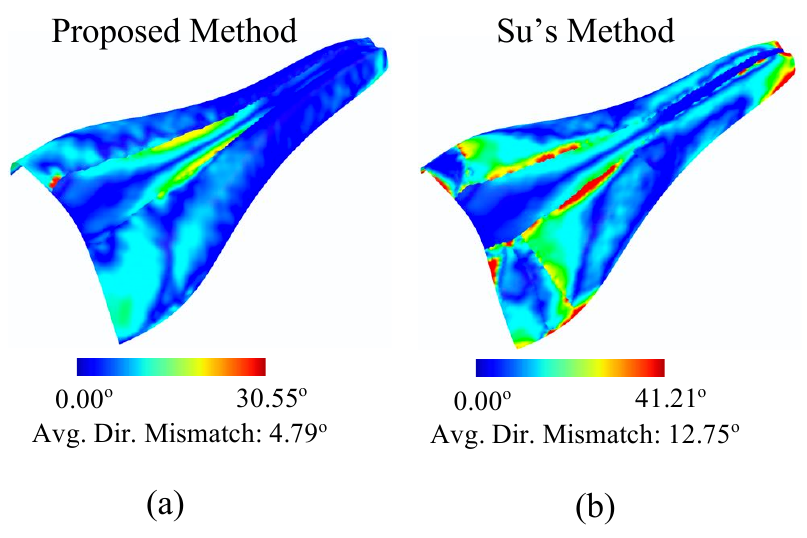}
	\caption{Direction alignment analysis of tool paths generated by the proposed method (a) and Su’s method (b).}
	\label{fig-seat-analysis}
\end{figure}

Case study 2 involved a bike seat surface (Fig.~\ref{fig-seat-result}a), which was also downloaded from GrabCAD. This surface represents a comprehensive example as it has convex, concave, and saddle regions. For the same reason, preferred feed directions generated for this surface exhibited non-smooth variations, necessitating surface segmentation. The segmented patches are shown with different colors in Figs.~\ref{fig-seat-result}b and \ref{fig-seat-result}c: three patches were identified for the proposed method, and four patches for Su's method (i.e., the separatrix-based method). No segmentation was needed for the classic iso-scallop method. The tool paths generated for individual patches are shown in Fig.~\ref{fig-seat-result}. A ball-end mill with radius 10mm was chosen, and the scallop height constraint was set as 0.5mm. The mismatch error analysis results are given in Fig.~\ref{fig-seat-analysis}.

\begin{figure*}[t]
	\centering
	\includegraphics[]{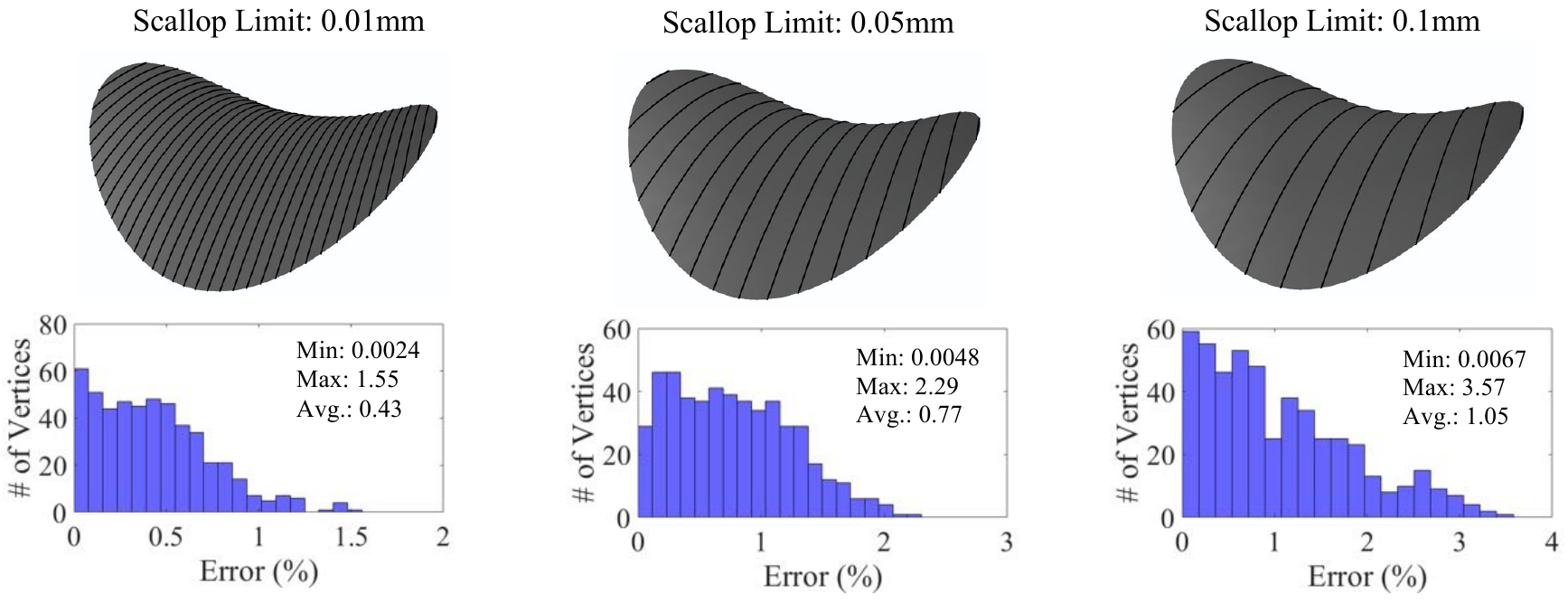}
	\caption{Error analysis of the proposed tool path planning method.}
	\label{fig-scallop-analysis}
\end{figure*}

Case study 3 analyzed the approximation error of the scallop height equation (\ref{eq-scallop-height-general}), using a saddle surface. This surface was chosen because it curves up in one direction and curves down in another direction, which allows us to analyze the comprehensive impact of varied curvatures (both the magnitude and the sign) on Eq.~\ref{eq-scallop-height-general}. Three scallop height constraints were used to generate tool paths (Fig.~\ref{fig-scallop-analysis}), with a flat-end mill of radius 2mm. (In typical finish machining, the tolerance often falls in $\left[0.01mm, 0.05mm\right]$, and the constraint 0.1mm can be viewed as an upper bound.) The error statistics shown in the figure were measured relative to the accurate calculation method presented in Ref.~\cite{li2004efficient}, and were based on 500 pairs of randomly sampled cutter contact points. It should be noted that we do not further provide the analysis results for other cutter types such as ball-end mills. This is because Eq.~\ref{eq-scallop-height-general} is identical to the classic formula (i.e., Eq.~\ref{eq-scallop-height-ball}) if ball-end mills are used, which indicates the best-case scenario. And the flat-end mill represents the worst-case scenario; other end mills fall in between.

\subsection{Discussion and limitations}
For the first two case studies, the overall length of the generated tool paths by the proposed method has been compared with those by the iso-parametric, iso-scallop \cite{feng2002constant}, preferred feed direction \cite{kumazawa2015preferred}, and enhanced preferred feed direction \cite{Su2020Initial} (the state of the art) methods. The results are summarized in Table~\ref{comparison-table}. In all comparisons, the proposed method is seen to generate the shortest tool paths. Large improvements are expected and confirmed for the classic iso-parametric and iso-scallop methods. Even for the state-of-the-art preferred feed direction method, the proposed method is still able to reduce the overall length by a notable percentage, above $7\%$, depending on the specific surface being considered.

The alignment analysis results shown in Figs.~\ref{fig-blade-analysis} and \ref{fig-seat-analysis} confirm the claim made at the beginning of this paper that reducing tool path length lies primarily in attaining the best tradeoff between the two requirements of following preferred feed directions and keeping the scallop height constant. Because the (enhanced) preferred feed direction method still focuses on iso-scallop tool paths, it failed to provide satisfactory tool path length reduction for the blade suction surface. One possible reason why the method did not work is: it uses a sequential tool path generation approach that leads to accumulated mismatch errors in the generated tool paths. On the other hand, the proposed method uses a global optimization (i.e., Eq.~\ref{equ-global-optimization}) to obtain all tool paths at once, and this can evenly distribute mismatch errors over the surface.

The error analysis results in Fig.~\ref{fig-scallop-analysis} show the effectiveness of Eq.~\ref{eq-scallop-height-general}. The maximum relative errors are all below $4\%$. Thus, if the required scallop height constraint is set as 0.01mm, the proposed method can generate tool paths with a worst-case scallop height constraint around $0.01 \pm 0.0004 \text{mm}$, which can provide satisfactory precision control in machining. However, this only holds for scallop height constraints in between 0.01mm and 0.1mm. When the constraint is made much larger, say 1mm, the approximation error is likely to have a significant increase. Fortunately, a scallop height constraint larger than 0.1mm is not commonly used in surface finish. 

By comparing the tool paths generated by the proposed method in the first two case studies, it is found that this method did not produce satisfactory tool paths near the borders between the segmented surface patches. In particular, there is no continuous, smooth transition across the borders. To solve this issue, additional constraints should be imposed on tool paths near the borders. Designing constraints to ensure continuity is not hard, but the challenge is to avoid affecting the interior tool paths' optimality. Such constraints remain unknown, and further development is required. As such, this issue can be considered as a serious limitation of the current work.

\begin{table*}[t]
	\renewcommand{\arraystretch}{1.25}
	\centering\small\setlength\tabcolsep{.225em}
	\centering
	\caption{Comparison of the proposed method with the various methods, using the iso-scallop results as references.}
	\begin{tabular}{c|c|c|c|c}
		\hline
		\multirow{2}{*}{Methods} & \multicolumn{2}{c}{Model 1: Turbine Blade}          & \multicolumn{2}{|c}{Model 2: Bike Seat} \\ \cline{2-5} 
								 & Path Length (mm) & \multicolumn{1}{c|}{Improvement} & Path Length (mm)      & Improvement     \\ \hline
		Iso-Parametric           &     4057.01             &             +9.10\%                     &       5405.56                &     +21.89\%            \\ \hline 
		Iso-Scallop \cite{feng2002constant}            &    3718.67              &            -                      &           4434.79            &       -          \\\hline
		\begin{tabular}[x]{@{}c@{}} Preferred Direction \\ Method \cite{kumazawa2015preferred}\end{tabular}   &     3693.11             &           -0.67\%                       &          4345.22             &       -2.02\%          \\\hline
		\begin{tabular}[x]{@{}c@{}}Enhanced Preferred \\ Direction Method \cite{Su2020Initial}\end{tabular}              &      3589.91            &               -3.46\%                   &          4297.28             &    -3.10\%             \\\hline
		Proposed                 &       3343.33           &          -10.09\%                        &       3857.68                &   -13.01\%   \\
		\hline           
		\end{tabular}
	 \label{comparison-table}
\end{table*}

\section{Conclusion}
\label{sec:conclusion}
A new method has been presented in this paper to generate length-optimal tool paths for freeform surface machining. The main features of this method include the minimum overall length of the generated tool paths and the simplicity of the formulation. These features are essentially achieved by (1) formulating the problem of minimizing tool path length as the problem of finding the closest satisfaction of constant scallop height and preferred feed directions, and (2) casting the closest satisfaction problem as a Poisson problem. The whole method consists primarily of two technical steps: (1) construct a vector field from a given preferred feed direction field and a constant scallop height constraint; and (2) find a scalar function whose gradient best approximates the vector field. New/improved methods have been presented to implement these two steps, and a series of case studies and comparisons have been conducted to validate the method.

Although the presented method is seen to be quite effective in the case studies conducted, there are a few limitations that should be noted here. During the implementation of the proposed method, it is found that, when the surface becomes very complex, the proposed method would segment the surface into many small patches, which could affect machining efficiency. Improving the surface segmentation algorithm is among the future research studies.

Another limitation is that the proposed method, in its current form, does not have a good treatment for transitioning tool paths from one segmented surface patch to its neighboring segmented surface patch. As a result, machining efficiency could be affected, and tool engagement/disengagement marks would be left on the machined surface. The method presented in Ref. \cite{Sun2017} may help but does not fit in our implicit tool path optimization framework. In the research to be carried out, the authors will focus on developing a new mechanism to carefully plan the tool paths across the borders so that a smooth tool path transition in those regions can be attained, and meanwhile the interior tool paths remain unchanged (or take the least change). 

It should also be noted that this work focuses on tool paths of minimum length, which may lead to seemingly sharp corners in the generated tool paths. Such non-smoothness in tool paths could affect the machining dynamics and consequently reduces machining efficiency. This states a serious limitation of the proposed method but also offers huge potential for improvement. Balancing the tool path length and smoothness has been touched upon in \ref{append-extend}. Further developing it, as well as including machining dynamics into tool path generation, can be very practically beneficial, which we would like to investigate in future work.

\section*{Acknowledgements}
This work has been funded by a UBC PhD Fellowship, a grant from Natural Sciences and Engineering Research Council of Canada (NSERC).

\appendix
\section{Extensibility of the proposed method}
\label{append-extend}
\noindent\textbf{Add smoothness to tool paths} \quad Tool path smoothness is favored in some applications like high speed machining. To add smoothness to tool paths generated by the optimization method in (\ref{equ-global-optimization}), we can keep the Laplacian of $\varphi$ small. (Laplacian of $\varphi$ is a widely used measure of smoothness \cite{Qiang2013iso}.) We thus optimize $\varphi$ by minimizing a linear combination of (\ref{equ-global-optimization}) and a smoothness energy:
\begin{equation}
	\min_{\varphi} \int_{\mathcal{S}} \left\|\nabla \varphi - V\right\|^2 + \lambda \left\|\Delta\varphi\right\|^2.
\end{equation}
where $\lambda$ is a weight term. This is a linear least squares problem.

\noindent\textbf{Lean towards preferred feed direction tool paths} \quad There are applications focusing primarily on preferred feed directions. For example, in manufacturing aero-engine blades, we want cutting passes to be consistent with fluid dynamics such that airflows can be guided by scallops on blades. For the present work to be applicable, we can merely minimize the energy in (\ref{equ-direction-align}), but the naive solution where $\varphi = \text{constant}$ must be avoided. To do so, we can forbid vanished gradients, and solve the following constrained optimization problem:
\begin{equation}
	\min_{\varphi} \int_{\mathcal{S}} \left\|D \cdot \nabla \varphi\right\|^2, \quad s.t. \quad  \int_{\mathcal{S}} \|\nabla \varphi\|^2 = 1.
\end{equation}
This constraint works because scaling $\nabla \varphi$ makes no difference for $\varphi$ to follow $D$. In the numerical setting, both the functional and the constraint become quadratic forms. Using Lagrange multipliers, solving the above problem amounts to finding the generalized eigenvector corresponding to the smallest generalized eigenvalue.

It should be noted that many existing work such as \cite{Sun2017, Su2020Initial} approached the above task through the following optimization model: $\min_{\varphi} \int_{\mathcal{S}} \left\| \nabla \varphi - D^{90^\circ}\right\|^2$. Unfortunately, this cannot give expected results. This energy encourages $\| \nabla \varphi \| = 1$ between adjacent streamlines, which pushes streamlines towards geodesic parallels. Real streamlines of a direction field are, however, far from geodesic parallels.

\noindent\textbf{Lean towards iso-scallop tool paths} \quad There are also applications where iso-scallop tool paths are mandatory, and preferred directions are secondary. For this purpose, we solve the following optimization problem:
\begin{equation}
	\min_{\varphi} \int_{\mathcal{S}} \left\|D \cdot \nabla \varphi\right\|^2, 
	\quad s.t. \quad \left\|\nabla \varphi\right\| = \sqrt{\frac{k_s+1/r_1}{8}}.
\end{equation}
Unlike the two previous optimization problems, this problem involves a complex hard constraint, necessitating the use of more sophisticated numerical optimization techniques. The augmented Lagrangian method that includes the constraint as a penalty term can be used to deal with this problem \cite{Nocedal06optimization}. In our previous work \cite{zou2014iso}, an optimization problem of similar complexity was involved, and the used solving procedures could be adapted to handle the problem here.

\bibliographystyle{elsarticle-num}
\bibliography{Bibliography}

\begin{thebibliography}{10}
\expandafter\ifx\csname url\endcsname\relax
  \def\url#1{\texttt{#1}}\fi
\expandafter\ifx\csname urlprefix\endcsname\relax\def\urlprefix{URL }\fi
\expandafter\ifx\csname href\endcsname\relax
  \def\href#1#2{#2} \def\path#1{#1}\fi

\bibitem{Altintas2005}
Y.~Altintas, C.~Brecher, M.~Weck, S.~Witt, Virtual machine tool, CIRP Annals -
  Manufacturing Technology 54 (2005) 115--138.

\bibitem{feng2002constant}
H.-Y. Feng, H.~Li, Constant scallop-height tool path generation for three-axis
  sculptured surface machining, Computer-Aided Design 34~(9) (2002) 647--654.

\bibitem{chiou2002machining}
C.-J. Chiou, Y.-S. Lee, A machining potential field approach to tool path
  generation for multi-axis sculptured surface machining, Computer-Aided Design
  34~(5) (2002) 357--371.

\bibitem{kumazawa2015preferred}
G.~H. Kumazawa, H.-Y. Feng, M.~J.~B. Fard, Preferred feed direction field: A
  new tool path generation method for efficient sculptured surface machining,
  Computer-Aided Design 67 (2015) 1--12.

\bibitem{crane2017heat}
K.~Crane, C.~Weischedel, M.~Wardetzky, The heat method for distance
  computation, Communications of the ACM 60~(11) (2017) 90--99.

\bibitem{zou2014iso}
Q.~Zou, J.~Zhang, B.~Deng, J.~Zhao, Iso-level tool path planning for free-form
  surfaces, Computer-Aided Design 53 (2014) 117--125.

\bibitem{lasemi2010recent}
A.~Lasemi, D.~Xue, P.~Gu, Recent development in cnc machining of freeform
  surfaces: A state-of-the-art review, Computer-Aided Design 42~(7) (2010)
  641--654.

\bibitem{suresh1994constant}
K.~Suresh, D.~Yang, Constant scallop-height machining of free-form surfaces,
  Journal of Engineering for Industry 116~(2) (1994) 253--259.

\bibitem{sarma1997geometry}
R.~Sarma, D.~Dutta, The geometry and generation of nc tool paths, Journal of
  Mechanical Design 119~(2) (1997) 253--258.

\bibitem{kim2007constant}
T.~Kim, Constant cusp height tool paths as geodesic parallels on an abstract
  riemannian manifold, Computer-Aided Design 39~(6) (2007) 477--489.

\bibitem{koren1996efficient}
Y.~Koren, R.~Lin, Efficient tool-path planning for machining free-form
  surfaces, ASME Journal of Engineering for Industry 118 (1996) 20--28.

\bibitem{tournier2002surface}
C.~Tournier, E.~Duc, A surface based approach for constant scallop
  heighttool-path generation, The International Journal of Advanced
  Manufacturing Technology 19~(5) (2002) 318--324.

\bibitem{Lee1998Nonisoparametric}
Y.-S. Lee, Non-isoparametric tool path planning by machining strip evaluation
  for 5-axis sculptured surface machining, Computer-Aided Design 30 (1998)
  559--570.

\bibitem{li2004efficient}
H.~Li, H.-Y. Feng, Efficient five-axis machining of free-form surfaces with
  constant scallop height tool paths, International Journal of Production
  Research 42~(12) (2004) 2403--2417.

\bibitem{wen2017cutter}
H.~Wen, J.~Gao, K.~Xiang, X.~Chen, Cutter location path generation through an
  improved algorithm for machining triangular mesh, Computer-Aided Design 87
  (2017) 29--40.

\bibitem{Chen2004Principle}
Z.~Chen, G.~Vickers, Z.~Dong, A new principle of cnc tool path planning for
  three-axis sculptured part machining—a steepest-ascending tool path, ASME
  Journal of Manufacturing Science and Engineering 126 (2004) 1--6.

\bibitem{anotaipaiboon2005tool}
W.~Anotaipaiboon, S.~S. Makhanov, Tool path generation for five-axis nc
  machining using adaptive space-filling curves, International Journal of
  Production Research 43~(8) (2005) 1643--1665.

\bibitem{moodleah2016five}
S.~Moodleah, E.~Bohez, S.~Makhanov, Five-axis machining of stl surfaces by
  adaptive curvilinear toolpaths, International Journal of Production Research
  54~(24) (2016) 7296--7329.

\bibitem{Sun2017}
Y.~Sun, S.~Sun, X.~J, G.~D, A unified method of generating tool path based on
  multiple vector fields for cnc machining of compound nurbs surfaces,
  Computer-Aided Design (2017).

\bibitem{ma2020toolpath}
J.-w. Ma, X.~Lu, G.-l. Li, Z.-w. Qu, F.-z. Qin, Toolpath topology design based
  on vector field of tool feeding direction in sub-regional processing for
  complex curved surface, Journal of Manufacturing Processes 52 (2020) 44--57.

\bibitem{kumazawa2012generating}
G.~H. Kumazawa, Generating efficient milling tool paths according to a
  preferred feed direction field, Master's thesis, University of British
  Columbia (2012).

\bibitem{liu2015tool}
X.~Liu, Y.~Li, S.~Ma, C.-h. Lee, A tool path generation method for freeform
  surface machining by introducing the tensor property of machining strip
  width, Computer-Aided Design 66 (2015) 1--13.

\bibitem{Su2020Initial}
C.~Su, X.~Jiang, G.~Huo, Y.~Sun, Z.~Zheng, Initial tool path selection of the
  iso-scallop method based on offset similarity analysis for global preferred
  feed directions matching, The International Journal of Advanced Manufacturing
  Technology 106~(7) (2020) 2675--2687.

\bibitem{lo1999efficient}
C.-C. Lo, Efficient cutter-path planning for five-axis surface machining with a
  flat-end cutter, Computer-Aided Design 31~(9) (1999) 557--566.

\bibitem{barakchi2010effective}
M.~J. Barakchi~Fard, H.-Y. Feng, Effective determination of feed direction and
  tool orientation in five-axis flat-end milling, ASME Journal of manufacturing
  science and engineering 132~(6) (2010).

\bibitem{Carmo76}
M.~P.~D. Carmo, Differential Geometry of Curves and Surfaces, Prentice-Hall,
  1976.

\bibitem{botsch2010polygon}
M.~Botsch, L.~Kobbelt, M.~Pauly, P.~Alliez, B.~L{\'e}vy, Polygon mesh
  processing, CRC Press, 2010.

\bibitem{kim2002toolpath}
T.~Kim, S.~E. Sarma, Toolpath generation along directions of maximum kinematic
  performance: a first cut at machine-optimal path, Computer-Aided Design
  34~(6) (2002) 453--468.

\bibitem{Crane2013DGP}
C.~Keenan, d.~G. Fernando, M.~Desbrun, P.~Schroder, Digital geometry processing
  with discrete exterior calculus, in: ACM SIGGRAPH 2013 courses, ACM, 2013.

\bibitem{belkin2003laplacian}
M.~Belkin, P.~Niyogi, Laplacian eigenmaps for dimensionality reduction and data
  representation, Neural computation 15~(6) (2003) 1373--1396.

\bibitem{jain2010data}
A.~K. Jain, Data clustering: 50 years beyond k-means, Pattern recognition
  letters 31~(8) (2010) 651--666.

\bibitem{Qiang2013iso}
Q.~Zou, J.~Zhao, Iso-parametric tool-path planning for point clouds,
  Computer-Aided Design 45~(11) (2013) 1459--1468.

\bibitem{Nocedal06optimization}
J.~Nocedal, S.~J. Wright, Numerical optimization, Springer, 2006.

\end{thebibliography}
\end{document}